\documentclass[%
 reprint,
 amsmath,amssymb,
 aps,pra
]{revtex4-1}

\usepackage{graphicx}
\usepackage{dcolumn}
\usepackage{bm}
\usepackage{epsfig}

\newcommand{\ket}[1]{\ensuremath{| #1 \rangle}}
\newcommand{\bra}[1]{\ensuremath{\langle #1 |}}
\newcommand{\ave}[1]{\ensuremath{\langle #1 \rangle}}

\begin{document}

\title{Phase coherent resonator detection as a complete quantum measurement of the two-mode spectral quantum state of light} 

\author{F. A. S. Barbosa$^1$, A. S. Coelho$^1$, K. N. Cassemiro$^2$, M. Martinelli$^1$, P. Nussenzveig$^1$, and A. S. Villar$^2$}

\email{villar@if.usp.br}

\affiliation{$^1$ Instituto de F\'\i sica, Universidade de S\~ao Paulo, P.O. Box 66318, 
05315-970 S\~ao Paulo, SP, Brazil \\ 
$^2$ Departamento de F\'\i{}sica, Universidade Federal de Pernambuco, 50670-901 Recife, PE, Brazil
}

\begin{abstract}
The introduction of phase coherence in the detection of quantum noise of light yields a pure quantum measurement of spectral modes. We theoretically show that such coherent quantum measurement performed with the technique of resonator detection (RD) is able to access any direction in the two-mode phase space of spectral sidebands under appropriate conditions, thus furnishing a complete measurement of the four-dimensional Wigner function. We obtain a realistic measurement operator for coherent RD by including the effects of imperfect resonator mode matching in our analysis.  
Moreover, we experimentally demonstrate the realization of phase coherent RD to characterize a two-mode displaced quantum state. 
\end{abstract}

\maketitle

\section{Introduction}

Measurement of the quantum noise of light is the main experimental tool to provide information on the quantum state of spectral modes in the continuous variables (CV) picture of quadrature observables. 
However, the usual procedure to access the spectral quantum noise provides neither a pure nor a complete quantum measurement of the two-mode spectral quantum state, due to lack of phase coherence in the measurement process~\cite{prl2013,pra2013}. 

Measurement mixedness currently restricts the faithful reconstruction of spectral quantum states to those presenting spectrally uniform energy distribution and Gaussian statistics, requiring the use of \textit{a priori} knowledge to achieve complete reconstruction. For this particular class of quantum states, it is possible to realize
 a pure quantum measurement of `effective' single-mode quadrature operators. Although such interpretation has been successfully utilized from the first experimental demonstrations of quantum noise squeezing up to the more recent observation of tripartite entanglement of spectral modes~\cite{slusherFirstSQZPRL85,heidmannPRL87,levenson4modesqzPRL87,oupereirakimblePRL92, villarPRL05,science09}, the experimental capability to unambiguously establish non-Gaussian features of the spectral quantum state has become of fundamental importance to move forward and consider more general quantum states of sideband modes~\cite{bartlett_prl02,bartlett_pra02,paris_pra31,ralphSidebandsSeparationPRA05,gauss2015}. 

The characterization and control of any unknown quantum state of spectral field modes requires the availability of a basic set of pure and complete quantum measurements. We have shown in a previous paper that the measurement technique of resonator detection (RD) is able to access novel aspects of each individual spectral mode even in the realistic non-ideal scenario of phase mixed quantum noise~\cite{prl2013}, by achieving an effective spatial separation of the sidebands~\cite{ralphSinglePhotonSidebandsPRA08}; moreover, RD is `complete' in the sense that it recovers all the information \textit{available} in this restricted scenario~\cite{pra2013}. In fact, the quantum noise of RD reveals additional information about the energy distribution in the two spectral sideband modes, a feature always missed by the widely employed measurement technique of spectral homodyne detection (HD). Even though the Gaussian character of the photocurrent statistics can be used to partially `undo' the incoherent effects introduced by phase mixing, \textit{a priori} knowledge about the Gaussian character of the quantum state is still assumed even in this favorable case~\cite{gauss2015}. 

In this paper, we theoretically show that extending RD to a phase coherent version furnishes a pure and complete measurement operator for the two spectral sideband modes, removing the need for \textit{a priori} assumptions about the quantum state. 
Our method fixes the incoherence gap in the downmixing chain by employing the eLO, as well as the optical LO, to produce the quantum state and thus ensure good phase coherence with both signals~\cite{naturebreitenbach}. In this experimental scenario, we employ RD to investigate a two-mode spectral quantum state produced by `phase modulation' of a laser beam. Finally, we investigate in theory the circumstances under which RD yields a novel pure and complete two-mode quantum measurement. 

These results generalize previous investigations where we have shown that RD achieves complete reconstruction of Gaussian quantum states in the usual incoherent detection scheme, a scenario requiring only four second-order quadrature moments to acquire complete information~\cite{prl2013,pra2013}. We now show RD to provide direct access to the four-dimensional phase space of two spectral modes, thus being able to reconstruct (assumption-free) any Wigner function by CV quantum state tomography~\cite{raymertomoPRL93,lvovsky09}. In the case of Gaussian quantum states, that would require ten moments to be accessed, although the technique would be more interesting when applied to non-Gaussian quantum states. Realistic limitations of the technique are investigated by explicitly considering in our analysis of the RD quantum observable the occurrence of spatial modal mismatch in the measurement setup. This measurement procedure should allow one to reconstruct arbitrary four-dimensional Wigner functions, in particular bringing the capability to unambiguously identify non-Gaussian features in the two-mode spectral quantum state, a currently scarce feature in the toolbox of spectral quantum noise measurement~\cite{paris_pra31,gauss2015}. 

We organize this paper as follows. In Sec.~\ref{sec:phasecoherentRD}, the basic theory is developed to show what should be expected from a complete two-mode measurement and how RD is able to achieve those requirements. We also investigate the effect of spatial mode mismatch in the measurement operator of RD. In Sec.~\ref{sec:experiment}, we realize the experimental implementation of coherent RD applied to the simplest two-mode quantum state possessing phase information: a coherent state produced by classical phase modulation. By keeping track of the spectral phase used in the electronic process of acquiring the Fourier components of the quantum noise, we are able to show phase sensitivity beyond what is currently attained in usual experiments. We offer our concluding remarks in Sec.~\ref{sec:conclusion}

\section{Phase coherent resonator detection}
\label{sec:phasecoherentRD}

The measurement technique of resonator detection is based on the dispersive property of an optical resonance~\cite{levenson,galatola,villarajp}. The quantum field of interest is first combined with the optical reference field (LO) and then reflected off an optical resonator; the total field is then detected and the beatnote signal analyzed. Due to the frequency-dependent phase shift and (crucially) modal attenuation, properties of individual spectral modes become accessible in the quantum noise. In particular, selective modal attenuation provides experimental access to the energy imbalance between spectral modes even in the phase mixing regime, a feature not recoverable with spectral HD~\cite{prl2013}. 

\subsection{Spectral photocurrent}

In the general context of spectral quantum noise measurements, the longitudinal modes of interest reside in the vicinity of the LO (with optical frequency $\omega_0$), 
separated from it by a beatnote frequency $\Omega\ll\omega_0$ selected at the detection step by the electronic reference. 
The upper sideband mode corresponds to the optical frequency $\omega = \omega_0+\Omega$ and the lower sideband  mode has the optical frequency $\omega = \omega_0-\Omega$. The LO is taken as an effective coherent state, with amplitude $\alpha = |\alpha|\exp(i\xi)$. The Fourier component $\hat I_\Omega$ of the photocurrent quantum noise at beatnote frequency $\Omega$ is described by the compact operator
\begin{equation}
\hat I_\Omega = e^{-i\xi}{|\alpha|}\,\hat a_{\omega_0-\Omega} + e^{i\xi}\,\hat a^\dag_{\omega_0+\Omega},
\label{eq:fotodetecaogeral}
\end{equation}
from which the two Hermitian observables corresponding to the cosine $\hat I_{\cos{}} = \frac12(\hat I_\Omega+\hat I_{-\Omega})$ and sine $\hat I_{\sin{}} = \frac{-i}{2}(\hat I_\Omega-\hat I_{-\Omega})$ photocurrent components can be determined.  For a field mode labelled by the optical frequency $\omega$, the photon annihilation $\hat a_{\omega}$ and creation $\hat a^\dag_{\omega}$ operators satisfy $[\hat a_\omega,\hat a^\dag_{\omega'}]= \delta(\omega-\omega')$. 
It is clear from Eq.~(\ref{eq:fotodetecaogeral}) that both spectral sideband modes contribute to the quantum noise, a consequence of the fact that the beatnote frequency $\Omega$ does not attain information on which optical signal has the largest frequency, whether the LO at $\omega_0$ or the sideband at $\omega$. 

The annihilation and creation operators in Eq.~(\ref{eq:fotodetecaogeral}) can be eliminated in favor of the amplitude $\hat p_\omega$ and phase $\hat q_\omega$ field quadratures, which obey the commutation relation $[\hat p_\omega,\hat q_{\omega'}]= 2i\delta(\omega-\omega')$, and read as $\hat p_\omega= \hat a_{\omega}+\hat a^\dag_{\omega}$ and $\hat q_\omega= -i(\hat a_{\omega}-\hat a^\dag_{\omega})$. This procedure reveals the symmetric ($\mathcal{S}$) and antisymmetric ($\mathcal{A}$) modal combinations of spectral modes as the `natural' modes of CV detection, represented by the quadrature observables~\cite{pra2013}
\begin{align}
\hat p_{s} = {\textstyle\frac{1}{\sqrt2}}\left(\hat p_\omega + \hat p_{\omega'}\right), & \qquad \hat p_{a} = {\textstyle\frac{1}{\sqrt2}}\left(\hat p_\omega - \hat p_{\omega'}\right), \nonumber\\
\hat q_{s} = {\textstyle\frac{1}{\sqrt2}}\left(\hat q_\omega + \hat q_{\omega'}\right), & \qquad \hat q_{a} =  {\textstyle\frac{1}{\sqrt2}}\left(\hat q_\omega - \hat q_{\omega'}\right).
\label{eq:symasymquadraturesdefinition}
\end{align}
In HD, the cosine photocurrent component measures only $\mathcal{S}$ modal quadratures, whilst the sine component yields access to the $\mathcal{A}$ mode. RD does not reveal a preferred modal basis. 

In most experiments, the Fourier amplitude of Eq.~(\ref{eq:fotodetecaogeral}) is probed only for its total energy, in which case phase information that could identify cosine or sine components is disregarded. Such approach is valid if one assumes Gaussian quantum states with uniform spectral energy distribution, in which case only second-order quadrature operator moments carry some interest. However, for the quantum measurement to be considere pure, phase coherence must exist between the spectral component of the quantum noise and the spectral quantum state: one requires the ability to coherently distinguish between the cosine and sine photocurrent components. We refer to this improved situation as \textit{phase coherent detection}.

\subsection{General form of a complete two-mode measurement}

The quantum state of a single-mode field can be represented on the phase space of CV quadrature observables. 
A complete single-mode quantum measurement must be able to determine any field quantum state $\hat \rho_\omega$. For instance, in the basis of eigenstates of $\hat q_\omega$, that would mean all matrix elements of the form $\bra{q_\omega}\hat \rho_\omega$\ket{q'_\omega}, where $q_\omega$ and $q'_\omega$ are eigenvalues of $\hat q_\omega$, must be accessible by measurement for the desired range of eigenvalues. 

In the case of photodetection, only the diagonal matrix elements are available to direct photodetection, a problem solved by measuring the same quantum state on many different bases (i.e. eigenstates of any combination of $\hat q_\omega$ and $\hat p_\omega$) by means of interferometric experiments such as HD. Borrowing measurement tools from standard techniques of quantum optics, quantum state tomography in phase space realizes precisely that. In fact, the favored point of view of CV quantum optics depicts the quantum state by quasi-probability distributions in phase space such as the Wigner function. 
For two-mode fields, the Wigner function exists in a four-dimensional phase space. Complete quantum state reconstruction requires the quantum measurement to access the probability distribution associated with any direction in the four-dimensional space. 

In the simpler case of a single-mode field (a subspace of the two-mode case), a complete measurement is required to deliver the family of local observables
\begin{equation}
\hat X_\omega(\varphi) = \cos\varphi\;\hat p_\omega + \sin\varphi\;\hat q_\omega. 
\label{eq:completesinglemode1}
\end{equation}
Each measurement operator $\hat X_\omega(\varphi)$ represents a direction of observation in the single-mode phase space controlled by the external parameter $\varphi$. 

Generalizing this idea, the reconstruction of the quantum state of a two-mode field comprised by optical modes $\omega$ and $\omega'$ requires additional access to the local observables of the single-mode field $\omega'$,  
\begin{equation}
\hat X_{\omega'}(\phi) = \cos\phi\;\hat p_{\omega'} + \sin\phi\;\hat q_{\omega'},
\label{eq:completesinglemode2}
\end{equation}
where the direction of observation in the second single-mode phase space is controlled by the independent phase parameter $\phi$. 
In addition to that, accessing two-mode coherences also requires the ability to perform a change of \textit{modal} basis, by coherently combining the two modes as in 
\begin{align}
\hat X_{\omega,\omega'}(\varphi,\phi,\theta) = & \cos\theta \;\hat X_\omega(\varphi) + \sin\theta\;\hat X_{\omega'}(\phi),
\label{eq:2modecombination}
\end{align}
where $\theta$ controls the relative contribution of modes $\omega$ and $\omega'$ to the measurement operator.
Thus a complete two-mode observable containing all the phase space projections needed to perform a complete two-mode measurement could be of the form
\begin{align}
\hat X_{\omega,\omega'}(\varphi,\phi,\theta) = & \cos\theta\cos\varphi\;\hat p_\omega +\cos\theta\sin\varphi\;\hat q_\omega \nonumber\\
 & + \sin\theta\cos\phi\;\hat p_{\omega'} + \sin\theta\sin\phi\;\hat q_{\omega'}.
\label{eq:general2modecompletemeasu}
\end{align}
The probability distribution associated with this observable provides information on the four-dimensional Wigner function of the two modes. From this expression, it is clear that any planar `slice' of the entire phase space could be accessed, since the phase space of any single mode composed by combining spectral modes $\omega$ and $\omega'$ would correspond to the observable $\hat X_{\omega,\omega'}(\varphi,\phi=\varphi,\theta)$. For instance, the $\mathcal{S}$ and $\mathcal{A}$ single-modes of HD could be realized by $\hat X_{\omega,\omega'}(\varphi,\varphi,\pi/4)$ and $\hat X_{\omega,\omega'}(\varphi,\varphi,-\pi/4)$, respectively,  while the observable $\hat X_{\omega,\omega'}(\varphi,\varphi+\pi/2,\theta)$ would access the type of two-mode correlations missed by HD, between any quadrature of $\mathcal{S}$ and $\mathcal{A}$ modes~\footnote{We note that the set of measurements available to spectral HD, even in the ideal case of phase coherent detection, is given by the family of observables $X_{\omega,\omega'}(\varphi,\varphi,\theta)$, i.e. limited by the condition $\varphi=\phi$ (i.e. the local phase spaces of sideband modes can be measured only in correlated directions), since a single controllable parameter, the LO phase, is available~\cite{prl2013,pra2013}. The technique is thus inherently limited in scope and incapable of accessing the complete two-mode phase space of spectral modes. }.

The observable of Eq.~(\ref{eq:general2modecompletemeasu}) transits seamlessly between single spectral modes and modal combinations delocalized in frequency. Hence two-mode quantum tomography in phase space could be in principle realized if this type of observable could be accessed in experiment~\cite{raymertomoPRL93,lvovsky09}. As we show below, RD provides just that. 


\subsection{Resonator detection measurement operator}

The field transformations taking place in RD are best described on the spectral modal basis. A high-finesse optical resonance centered at frequency $\omega_c$ with bandwidth $2\gamma$ performs in reflecting each spectral field mode the quantum operation 
\begin{equation}
\hat a_\omega^{out} = r(\Delta_\omega)\,\hat a_\omega + t(\Delta_\omega)\,\hat b_\omega,
\label{eq:inputoutput}
\end{equation}
where $\hat a_\omega$ is the annihilation operation of the field of interest impinging on the resonator and $\hat b_\omega$ represents the modes in vacuum state transmitted by the resonator. The reflection $r(\Delta_\omega)$ and transmission $t(\Delta_\omega)$ coefficients are functions of the dimensionless detuning $\Delta_\omega = (\omega - \omega_c)/\gamma$. The measured output field $\hat a_\omega^{out}$ is hence a coherent combination of the spectral mode of interest upon reflection and leaked vacuum, after the first is attenuated and phase shifted by the amount
\begin{equation}
r(\Delta_\omega) \approx -\frac{\sqrt{d}+i\Delta_\omega}{1-i\Delta_\omega} = \sqrt{1-T(\Delta_\omega)}\,e^{i\Psi(\Delta_\omega)}.
\label{eq:rresonator}
\end{equation}
This expression holds in the high finesse limit, where $d$, the \textit{impedance matching parameter} (a fixed property of the optical resonator), represents the fraction of light intensity reflected at exact resonance. For an impedance-matched resonator, light is completely transmitted at resonance (i.e. $d=0$ and $|r(0)|=0$), whereas for a lossless resonator light is totally reflected (i.e. $d=1$ and $|r(\Delta_\omega)|=1, \forall \Delta_\omega$). The spectral attenuation $T(\Delta_\omega)$ and phase shift $\Psi(\Delta_\omega)$ follow the explicit expressions
\begin{align}
T(\Delta_\omega) & = \frac{1-d}{1+\Delta_\omega^2}, \\
\Psi(\Delta_\omega) & = \arctan\left(\frac{\Delta_\omega-1}{\Delta_\omega+1}\,\frac{\Delta_\omega-\sqrt{d}}{\Delta_\omega+\sqrt{d}}\right). 
\end{align}

Quantum state reconstruction requires scanning the optical resonance to sequentially transform according to Eq.~(\ref{eq:inputoutput}) the lower sideband, LO, and the upper sideband just prior to photodetection. The simplest transformation occurs for the LO mode: its mean amplitude $\alpha$ is attenuated and phase-shifted as $\alpha^{out} = r(\Delta)\alpha$, where $\Delta= (\omega_0-\omega_c)/\gamma$ is the detuning between LO and optical cavity. It is convenient to label the other optical frequencies with respect to the LO mode, since it is not only the phase reference but also the frequency ruler that defines the sideband modes. We write $\Delta_\omega$ as a function of the LO detuning according to $\Delta_\omega=\Delta+\Omega/\gamma$, where the sideband mode $\omega$ is labeled relative to LO by $\Omega=\omega-\omega_0$, $\Omega\ll\omega_0$. Applying these considerations to the transformations of Eq.~(\ref{eq:inputoutput}), the photodetection operator of Eq.~(\ref{eq:fotodetecaogeral}) yields the spectral photocurrent operator of RD as 
\begin{equation}
\hat J_{\Omega}(\Delta) = R^*_{\Omega}(\Delta)\,\hat a_\Omega + R_{-\Omega}(\Delta)\,\hat a^\dag_{-\Omega}+\hat J_v,
\label{eq:Jomegadef}
\end{equation}
where the notation has been simplified in $\hat a_{\omega_0\pm\Omega}\rightarrow\hat a_{\pm\Omega}$, $\hat J_v$ stands for vacuum modes, and the transformation coefficients are
\begin{align}
R_{\Omega}(\Delta) & = e^{i\Psi(\Delta)}\,r^*(\Delta+\Omega/\gamma)\nonumber\\
& = e^{i\Psi(\Delta)}\sqrt{1-T(\Delta +\Omega/\gamma)} \,e^{-i\Psi(\Delta +\Omega/\gamma)}.
\end{align}

The three distinct spectral regions of the field (lower sideband, LO, and upper sideband) are simultaneously transformed by $R_{\Omega}(\Delta)$ as the cavity resonance frequency is scanned during quantum state reconstruction. Let us suppose for the sake of the argument that sideband modes are separated by a frequency interval $\Omega\gg\gamma$ from the LO (narrow band resonator), so that the transformations affecting each region of detuning do not interfere with one another. In this case, we can separate the effect of the resonator on the reflected field according to three distinct situations: in \textit{(i)} and \textit{(iii)}, either upper or lower sideband is nearly resonant with the optical resonator; in \textit{(ii)}, the LO undergoes the sole action of the resonator. We now detail those three possible field transformations as quantum operations affecting the measurement observable associated with the two-mode quantum state.

In region \textit{(i)}, the optical resonator is nearly resonant with the upper sideband ($\Delta\approx-\Omega/\gamma$) and the remaining terms of Eq.~(\ref{eq:Jomegadef}) can be considered as off-resonant, i.e. $e^{i\Psi(\Delta)}\approx 1$ and $|r(\Delta-\Omega/\gamma)|\approx 1$. The spectral photocurrent operator then simplifies to 
\begin{align}
\hat J_{\Omega}(\Delta) \approx & \sqrt{1-T(\Delta+\Omega/\gamma)}\,e^{i\Psi(\Delta+\Omega/\gamma)}\,\hat a_\Omega  + \hat a^\dag_{-\Omega} \nonumber\\
& + \sqrt{T(\Delta+\Omega/\gamma)}\,\hat b_\Omega.
\label{eq:jomegaregion1}
\end{align}
The resonator detuning controls in this case the partial substitution of the upper sideband mode by the vaccum field and at the same time the \textit{rotation} of its quasi-probability distribution by the angle $\Psi(\Delta+\Omega/\gamma)$~\cite{huntingtonPRA05}~\footnote{We note that the vacuum annihilation operator $\hat b_\Omega$ has been simplified in Eq.~(\ref{eq:jomegaregion1}) to incorporate phase shifts that do not affect the quantum noise statistics (we employ this simplifying procedure wherever possible).}. 

Spectral region \textit{(ii)} sees the LO field interacting with the optical resonator ($\Delta\approx0$), so that the off-resonant terms of Eq.~(\ref{eq:Jomegadef}) now refer to the optical sidebands, which remain nearly unaffected as $r(\Delta\pm\Omega/\gamma)\approx1$. Under these conditions, the photocurrent operator reads as
\begin{align}
\hat J_{\Omega}(\Delta) \approx & \sqrt{1-T(\Delta)}\left(e^{-i\Psi(\Delta)}\,\hat a_\Omega + e^{i\Psi(\Delta)}\,\hat a^\dag_{-\Omega}\right).
\label{eq:jomegaregion2}
\end{align}
Apart from the attenuation factor ${1-T(\Delta)}$, this observable essentially reproduces the measurement operator of spectral HD. The attenuation changes the absolute power of the spectral noise, resulting in a redefinition of the standard quantum level (SQL) of the shot noise and eventually decreasing the quality of the quantum signal in face of technical noise. However, as a quantum measurement, the phase space rotation induced by the LO phase delay in region \textit{(ii)} is equivalent to a change of the measurement operator to the $\mathcal{S}/\mathcal{A}$ modal basis. In this sense, RD \textit{contains} HD as part of the modal transformations leading to the measurement operator. 

Finally, spectral region \textit{(iii)} applies the same transformation of Eq.~(\ref{eq:jomegaregion1}) to the lower sideband ($\Delta\approx\Omega/\gamma$). Explicitly, the spectral photocurrent operator in this region reads as 
\begin{align}
\hat J_{\Omega}(\Delta) \approx & \; \hat a_\Omega + \sqrt{1-T(\Delta-\Omega/\gamma)}\,e^{-i\Psi(\Delta-\Omega/\gamma)}\,\hat a^\dag_{-\Omega} \nonumber\\
& + \sqrt{T(\Delta-\Omega/\gamma)}\,\hat b^\dag_{-\Omega}.
\label{eq:jomegaregion3}
\end{align}
The lower sideband mode undergoes transformations equivalent to those experienced by the upper sideband in region 1 as the detuning is varied. 

Since the measurement operator in region \textit{(ii)} essentially mimics HD, the novel features of RD must be present in regions \textit{(i)} and \textit{(iii)}. The exact effect on quantum noise of the special quantum transformations in those regions strongly depends on the impedance matching parameter $d$. Resonator detection has two independent scenarios of interest, determined by the extreme values of $d$. In the first scenario, a lossless ideal resonator ($d=1$) will have the sole effect of dephasing the spectral modes, i.e. no modal attenuation occurs. The spectral photocurrent operator then takes essentially the same form in regions \textit{(i)}, \textit{(ii)} and \textit{(iii)}. For instance, in region~\textit{(ii)} the operator becomes
\begin{align}
\hat J^{(1)}_{\Omega}(\Delta) \approx & \; e^{i\Psi(\Delta+\Omega/\gamma)/2}\left(e^{i\Psi(\Delta+\Omega/\gamma)/2}\hat a_\Omega  \right. \nonumber \\
& \left. + e^{-i\Psi(\Delta+\Omega/\gamma)/2}\hat a^\dag_{-\Omega}\right),
\end{align}
which has the same form of Eq.~(\ref{eq:jomegaregion2}) although with the opposite direction of rotation in phase space (the leading phase has no effect in measurement results). The same expression is valid for region~\textit{(iii)} by flipping the sign of the leading phase. 
Hence the three detuning regions implied by the lossless resonator do not differ from one another in the form of the measurement operator: in this case, RD becomes solely based on modal phase shifts and thus completely equivalent to spectral HD. This fact sheds some light on the reasons why HD is incomplete as a two-mode measurement. It is essentially bound to measure the spectral modes indistinguishably, implying that only symmetric or antisymmetric modal combinations can be accessible by principle. 

Novel features regarding quantum state reconstruction appear in the second extreme scenario. The impedance-matched resonator ($d=0$) is built as to promote the complete exchange of reflected and transmitted modes at exact resonance. Then in regions \textit{(i)} and \textit{(iii)} the reflected field furnishes an effective measurement of individual sideband modes, in the same spirit of Eq.~(\ref{eq:completesinglemode1}). In region \textit{(i)}, for the special value of detuning $\Delta = -\Omega/\gamma$, one has $T(\Delta+\Omega/\gamma) =1$, and Eq.~(\ref{eq:jomegaregion1}) yields  
\begin{align}
\hat J_{\Omega}(-\Omega/\gamma) = \hat a^\dag_{-\Omega} + \hat b_\Omega,
\end{align}
i.e. a measurement of the lower sideband mode (contaminated by vacuum noise). A similar relation is obtained for Eq.~(\ref{eq:jomegaregion3}) at the detuning $\Delta = \Omega/\gamma$, and a direct measurement of upper sideband mode follows. Thus RD provides the ability to change the measurement modal basis. 
In region \textit{(ii)}, the effect of modal exchange comes without conceptual consequences, as attenuation of the LO mode simply decreases the LO amplification of sideband quantum noise. LO attenuation implies a recalibration of the SQL. 
Total attenuation of the LO, however, could produce technical problems due to a pathological measurement in the ideal case $d=0$: Technical noise would dominate the spectral photocurrent near the particular detuning $\Delta\approx0$. In reality, this effect limits the minimum value of $d$ attainable in experiment by considering the smallest intensity at exact resonance still producing quantum noise with the desired signal-to-noise ratio. Alternatively, the input LO intensity could in principle be increased accordingly in this detuning region to compensate the attenuation in reflection. We finally note that vacuum contamination of the field mode is a common form of technical noise in the tomographic reconstruction of quantum states; in fact, every type of modal contamination occurring in a realistic experiment leads to a decrease of signal-to-noise ratio. It is only necessary to calibrate the technical noise and deconvolute the measured probability distribution. 


\subsection{Phase coherent RD as a complete quantum measurement}

The actual observables associated with the RD quantum measurement are the photocurrent components $\hat J_{\cos{}}$ and $\hat J_{\sin{}}$ defined by the identity $\hat J_{\Omega}=(\hat J_{\cos{}}+i\hat J_{\sin{}})/\sqrt{2}$. To be accessed, a well defined phase relation between the optical LO and the eLO reference employed to extract the photocurrent Fourier $\Omega$ component is required. In most experiments, incoherence in the spectral analysis leads to the spectrum noise power $S(\Omega) = \ave{\hat J_{\Omega}\hat J_{-\Omega}}$ as the sole meaningful quantity amenable to measurement~\cite{gauss2015}. 

Phase coherent detection brings one additional controllable parameter to the phase space measurement. Any combination of the spectral observables $\hat J_{\cos{}}$ and $\hat J_{\sin{}}$ becomes available by tuning the relative phase $\Theta$ between the optical LO and the eLO. Phase coherence implies that the spectral component of Eq.~(\ref{eq:fotodetecaogeral}) could also be chosen as $\hat I_\Omega \rightarrow e^{i\Theta}\hat I_\Omega$, where $\Theta$ becomes a controllable parameter. The form of the general quantum measurement of phase coherent resonator detection is hence $\hat J_{\Theta} = \cos\Theta\,\hat J_{\cos{}} + \sin\Theta\,\hat J_{\sin{}}$.

We now analyze the RD measurement operator of Eq.~(\ref{eq:Jomegadef}) to show that it is able to realize the general two-mode measurement of Eq.~(\ref{eq:general2modecompletemeasu}) under certain conditions. 
In doing so, we substitute in Eq.~(\ref{eq:Jomegadef}) the field annihilation and creation operators in favor of the quadrature observables. We either choose the spectral ($\hat p_{\pm\Omega}$ and $\hat q_{\pm\Omega}$) or the $\mathcal{S}/\mathcal{A}$ modal quadrature operators defined in Eq.~(\ref{eq:symasymquadraturesdefinition}) as convenient. 
We consider at first the scenario of a narrow linewidth resonator ($\gamma\ll\Omega$) for the sake of clarity. 

Starting with the ideal lossless resonator ($d=1$), for which $|R_{\Omega}(\Delta)|=1$, the observables $\hat J_{\cos{}}$ and $\hat J_{\sin{}}$ establish in this case that $\mathcal{S}$ and $\mathcal{A}$ modes form a privileged modal basis of measurement in all three detuning regions, since Eq.~(\ref{eq:Jomegadef}) yields simply 
\begin{align}
\label{eq:privSAcos}
{\textstyle\frac{1}{|\alpha^{out}|}}\hat J_{\cos{}} & \approx \cos\Psi\,\hat p_s + \sin\Psi\,\hat q_s, \\
\label{eq:privSAsin}
{\textstyle\frac{1}{|\alpha^{out}|}}\hat J_{\sin{}} & \approx -\sin\Psi\,\hat p_a + \cos\Psi\,\hat q_a,
\end{align}
where we have multiplied the measurement operators by the inverse of the reflected LO power $|\alpha^{out}|$ to better compare the detuning regions (we note that the SQZ is proportional to $|\alpha^{out}|^2$), and disregarded leading phases. 
While the argument of $\Psi$ vary among the detuning regions, the general \textit{form} of the measurement operators always favor the $\mathcal{S}$ and $\mathcal{A}$ modal basis. 
Although these two modes can be individually measured, they are are both rotated in phase space by one and the same parameter $\Psi$. The most general measurement operator $\hat J_{\Theta}$ coherently combines $\hat J_{\cos{}}$ and $\hat J_{\sin{}}$, as shown in Eq.~(\ref{eq:2modecombination}). 
The general observable assumes the same form in the three detuning regions, 
\begin{align}
\frac{1}{|\alpha^{out}|}\hat J_{\Theta} & \approx \cos\Theta\left(\cos\Psi\,\hat p_s + \sin\Psi\,\hat q_s\right) \nonumber\\
 & +\sin\Theta\left(-\sin\Psi\,\hat p_a + \cos\Psi\,\hat q_a\right),
 \label{eq:operadormedidaregiao1}
\end{align}
Performing a change of modal basis, in terms of spectral sideband quadratures this expression reads as
\begin{align}
\frac{1}{|\alpha^{out}|}\hat J_{\Theta} & \approx \cos(\Theta+\Psi)\,\hat p_\Omega + \sin(\Theta+\Psi)\,\hat q_\Omega \nonumber\\
 & +\cos(\Theta-\Psi)\,\hat p_{-\Omega} - \sin(\Theta-\Psi)\,\hat q_{-\Omega},
\end{align}
from which it becomes clear, by comparison with Eq.~(\ref{eq:general2modecompletemeasu}), that RD without modal attenuation can not attain a complete two-mode quantum measurement. The missing sectors of the four-dimensional phase space are in this case the same as in HD, and correspond to the energy asymmetry of spectral sideband modes in the case of Gaussian states or, on the $\mathcal{S}$ and $\mathcal{A}$ modal basis, to certain modal correlations involving the same direction of observation in the two-mode phase space~\cite{prl2013}. 

The second extreme working scenario of RD corresponds to the ideal impedance-matched resonator ($d=0$), for which the mode at exact resonance is completely replaced in reflection by a mode in the vacuum state. 
In detuning region \textit{(ii)} ($\Delta\approx 0$), 
the LO phase shift continues to favor the $\mathcal{S}$ and $\mathcal{A}$ modal basis. Hence the operator of  Eq.~(\ref{eq:operadormedidaregiao1}) also describes the quantum measurement in this case. 
RD with the impedance-matched resonator accesses in region \textit{(ii)} the single-mode quantum state of $\mathcal{A}$ and $\mathcal{S}$ modal basis and correlations between them exactly as in HD. 

In region \textit{(i)}, the resonator attenuates and phase-shifts only the upper sideband mode, thus breaking the symmetry between sideand modes in the quantum measurement. Eq.~(\ref{eq:jomegaregion1}) yields the photocurrent observables 
\begin{align}
\label{eq:Jcosquadregion1}
\hat J_{\cos{}} \approx & \sqrt{1-T_{\Omega}}\left(\cos\Psi_{\Omega}\,\hat p_\Omega - \sin\Psi_{\Omega}\,\hat q_\Omega \right)+ \hat p_{-\Omega}\nonumber\\
& + \sqrt{T_{\Omega}}\,\hat u,\\
\label{eq:Jsinquadregion1}
\hat J_{\sin{}} \approx & \sqrt{1-T_{\Omega}}\left(\sin\Psi_{\Omega}\,\hat p_\Omega + \cos\Psi_{\Omega}\,\hat q_\Omega \right) - \hat q_{-\Omega} \nonumber\\
& + \sqrt{T_{\Omega}}\,\hat v,
\end{align}
where $\hat u$ and $\hat v$ represent orthogonal quadratures of a mode in the vacuum state and notation has been simplified to $T_{\Omega} = T(\Delta+\Omega/\gamma)$ and $\Psi_{\Omega} = \Psi(\Delta+\Omega/\gamma)$. 
We note that as expected the $\mathcal{S}/\mathcal{A}$ modal basis becomes 
again a convenient measurement basis in case the resonator is far off resonance, since $\Psi(\Delta\ll-\Omega/\gamma)\rightarrow0$. But of particular interest is the situation of resonant upper sideband ($\Delta = -\Omega/\gamma$), in which the two spectral photocurrent components perform a simultaneous measurement of lower sideband conjugate quadratures, since  
\begin{align}
\hat J_{\cos{}} = \hat p_{-\Omega} + \hat u \quad\text{and} \quad \hat J_{\sin{}}= -\hat q_{-\Omega}+\hat v. 
\end{align}
The added vacuum noise ensures the 3~dB noise penalty typical of simultaneous measurements of non-commuting observables~\cite{royerWignermeasurement_fp89,3dbbellj}. Phase coherent detection allows any direction of observation in the phase space of lower sideband when the resonator is perfectly tuned to the upper sideband, since in this case one can construct the general measurement operator 
\begin{align}
\hat J_{\Theta} = \cos\Theta\,\hat p_{-\Omega} - \sin\Theta\,\hat q_{-\Omega} + \hat u',
\label{eq:lowersidebandmeas}
\end{align}
where the single-mode vacuum mode operator has been redefined as $\hat u'$ by a convenient change of modal basis. This measurement operator has the form of the single-mode observable of Eq.~(\ref{eq:completesinglemode1}) needed as part of a complete two-mode measurement. RD provides quantum measurements of the two-mode field in different modal basis as the resonator detuning is varied between regions. 

In general, the measurement operator in region~\textit{(i)} [Eq.~(\ref{eq:lowersidebandmeas})] produces an arbitrary change of modal basis. In fact, the observables of Eqs.~(\ref{eq:Jcosquadregion1}) and~(\ref{eq:Jsinquadregion1}) can be written in the form 
\begin{align}
\label{eq:Jcosquadregion1simples}
\hat J_{\cos{}} \approx & \;\hat p_{-\Omega} + \cos\xi\,\hat X_\Omega(\Psi) + \sin\xi\,\hat u,\\
\label{eq:Jsinquadregion1simples}
\hat J_{\sin{}} \approx & \;-\hat q_{-\Omega} + \cos\xi\,\hat X_\Omega(\Psi-\pi/2) + \sin\xi\,\hat v,
\end{align}
where $\cos\xi = \sqrt{1-T_\Omega}$ 
($0\leq\xi\leq\pi/2$) and $\hat X_\Omega(\Psi)=\cos\Psi_{\Omega}\,\hat p_\Omega - \sin\Psi_{\Omega}\,\hat q_\Omega$ is the generalized quadrature of mode $\Omega$. The observables above represent a continuous change of modal basis dependent on the parameter $\xi$ as the upper sideband is attenuated close to resonance (in fact, $\xi\rightarrow0$ for $\Delta\ll-\Omega/\gamma$ and  $-\Omega/\gamma\ll\Delta\ll0$). One may denote the new quadrature basis as $\hat P_\xi = (\hat p_{-\Omega} + \cos\xi\,\hat X_\Omega(\Psi))/\sin\xi$ and $\hat Q_\xi = (-\hat q_{-\Omega} + \cos\xi\,\hat X_\Omega(\Psi-\pi/2))/\sin\xi$. Phase coherent detection provides for each modal transformation above (fixed $\xi$) the possibility of observing any direction in its phase space, since the combination of cosine and sine photocurrent components yields in this case
\begin{align}
\hat J_{\Theta} \approx & \cos\Theta\,\sin\xi\,\hat P_\xi + \sin\Theta\,\sin\xi\,\hat Q_{\xi+\pi/2} + \sin\xi\,\hat u'. 
\end{align}
Hence region~\textit{(i)} sees the continuous transformation of measurement basis from $\mathcal{S}$ and $\mathcal{A}$ modes ($\hat P_\xi\approx\hat p_s$ and $\hat Q_\xi\approx\hat q_a$ for $\Delta \ll- \Omega/\gamma$) to a direct measurement of lower sideband phase space at $\Delta = -\Omega/\gamma$ [Eq.~(\ref{eq:lowersidebandmeas})] and back to $\mathcal{S}$ and $\mathcal{A}$ modes in region~\textit{(ii)} ($\Delta \approx 0$). 
Similar considerations lead to the expressions of quantum measurements in region \textit{(iii)}, with the upper sideband assuming the same prominent role that the lower sideband has in region~\textit{(i)}. The quantum observables are obtained from Eqs.~(\ref{eq:Jcosquadregion1simples}) and~(\ref{eq:Jsinquadregion1simples}) by the exchange $\Omega\rightarrow-\Omega$. In particular, the phase space of upper sideband is measured at exact resonance with the lower sideband at $\Delta=\Omega/\gamma$, yielding a quantum measurement operator analogous to that of Eq.~(\ref{eq:lowersidebandmeas}).

Thus resonator detection employing the impedance-matched narrow band resonator in principle accesses the two-mode quantum state by measuring it in two different single-mode basis: $\{\hat p_\Omega,\hat q_\Omega\}$ in region \textit{(i)}, $\{\hat p_s,\hat q_s\}$ and $\{\hat p_a,\hat q_a\}$ in region \textit{(ii)}, and $\{\hat p_{-\Omega},\hat q_{-\Omega}\}$ in region \textit{(iii)}. Intermediate modal combinations are also available, according to Eqs.~(\ref{eq:Jcosquadregion1simples}) and~(\ref{eq:Jsinquadregion1simples}). 
The ability to observe single-mode phase spaces for different sets of modal basis in the two-mode space of sidebands allows RD to access the complete four-dimensional phase space of the Wigner function. 


\subsection{Two-mode quantum state reconstruction}

The three detuning regions where the RD measurement operator presents different well defined behaviors in the simplifying scenario of a narrow linewidth resonator merge seamlessly in the actual measurement with a typical resonator, for which the condition $\gamma\ll\Omega$ is not necessarily satisfied. The complete expression for the general quantum observable of resonator detection with the impedance matched resonator is 
\begin{align}
\hat J_{\Theta}(\Delta) = 
&\; \cos\xi_{\Omega} \,\hat X_{\Omega}(\theta) + \cos\xi_{-\Omega} \,\hat X_{-\Omega}(\theta') \nonumber\\
& + \sin\xi_{\Omega} \,\hat v_{\Omega} + \sin\xi_{-\Omega} \,\hat v_{-\Omega},
\end{align}
where $\theta = \Psi_0-\Psi_{\Omega}+\Theta$ and $\theta'= \Psi_0-\Psi_{-\Omega}-\Theta+\pi$.
This measurement operator has the form of the complete two-mode operator of Eq.~(\ref{eq:general2modecompletemeasu}), although combined with additional vacuum contributions that preserve the commutation relations when substituting individual quantum modes by vacuum fields. 

\begin{figure}[t]
\epsfig{width=\columnwidth,file=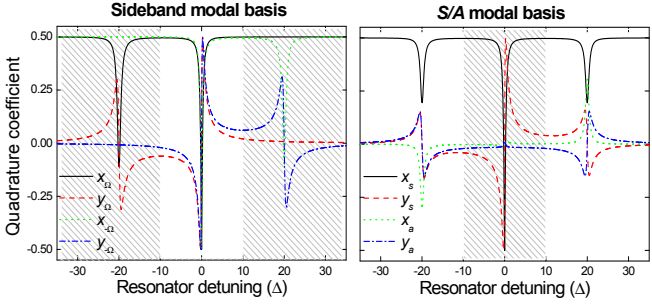}
\caption{Coefficients of resonator detection as functions of resonator detuning $\Delta$. Sideband frequency is $\Omega=20\gamma$. Left: Coefficients as written on the modal basis of spectral sidebands [Eqs.~(\ref{eq:Jcosxy})--(\ref{eq:Jsinxy})]. The dashed regions show the phase space rotation of individual sideband modes. Right: Coefficients as written on the $\mathcal{S}$ and $\mathcal{A}$ modal basis [Eqs.~(\ref{eq:JcosxyAS})--(\ref{eq:JsinxyAS})]. Phase space rotation of $\mathcal{S}$ and $\mathcal{A}$ modes occurs on the dashed region.}
\label{fig:coefs}
\end{figure}

Alternatively, the observables of RD can also be written in terms of real functions of detuning defined by $R_{\pm\Omega}(\Delta) = x_{\pm\Omega}(\Delta)+iy_{\pm\Omega}(\Delta)$, 
\begin{align}
\label{eq:Jcosxy}
\hat J_{\cos{}} & = x_{\Omega}\,\hat p_{\Omega} + y_{\Omega}\,\hat q_{\Omega} + x_{-\Omega}\,\hat p_{-\Omega} + y_{-\Omega}\,\hat q_{-\Omega} + \hat w_\mathrm{cos},\\
\label{eq:Jsinxy}
\hat J_{\sin{}} & = -y_{\Omega}\,\hat p_{\Omega} + x_{\Omega}\,\hat q_{\Omega} + y_{-\Omega}\,\hat p_{-\Omega} - x_{-\Omega}\,\hat q_{-\Omega} + \hat w_\mathrm{sin},
\end{align}
a more convenient expression to perform numerical fitting to the experimental data. The vacuum operators are defined as $\hat w_{\cos{}} = \sqrt{T_\Omega}\;\hat u_\Omega + \sqrt{T_{-\Omega}}\;\hat u_{-\Omega}$ and $\hat w_{\sin{}} = \sqrt{T_\Omega}\;\hat v_\Omega + \sqrt{T_{-\Omega}}\;\hat v_{-\Omega}$. In the basis of $\mathcal{S}$ and $\mathcal{A}$ modes, these observables read as 
\begin{align}
\label{eq:JcosxyAS}
\hat J_{\cos{}} & = x_{s}\,\hat p_{s} + y_{s}\,\hat q_{s} + x_{a}\,\hat p_{a} + y_{a}\,\hat q_{a} + \hat w_{\cos{}},\\
\label{eq:JsinxyAS}
\hat J_{\sin{}} & = -y_{s}\,\hat p_{s} + x_{s}\,\hat q_{s} + y_{a}\,\hat p_{a} - x_{a}\,\hat q_{a} + \hat w_{\sin{}}.
\end{align}

Figure~\ref{fig:coefs} depicts the coefficients of quadrature operators in Eqs.~(\ref{eq:Jcosxy})--(\ref{eq:Jsinxy}) for a narrow band resonator. The main features of resonator detection can be seen in the three different detuning regions separated in the figure by shaded backgrounds. The coefficients appearing in $\hat J_{\cos{}}$ and $\hat J_{\sin{}}$ are shown on two different modal basis as functions of cavity detuning. 
The spectral basis of sideband modes (top row) is better suited to understand the quantum transformations in regions~\textit{(i)} ($\Delta<-10$) and~~\textit{(iii)} ($\Delta>10$), where the individual sidebands undergo phase shift and attenuation. The $\mathcal{S}$ and $\mathcal{A}$ modal basis (bottom row) simplifies the description of region~~\textit{(ii)} ($-10<\Delta<10$), whereby it is clear that the cosine component promotes a phase space rotation of the $\mathcal{S}$ mode, while the sine component rotates the phase space of the $\mathcal{A}$ mode.

\subsection{Modal contamination}

\begin{figure}[ht]
\epsfig{width=\columnwidth,file=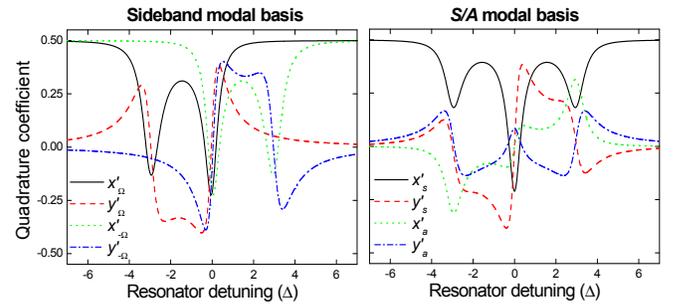}
\caption{Coefficients of resonator detection with the measurement parameters of Sec.~\ref{sec:experiment}. Line styles follow the same convention as in Fig.~\ref{fig:coefs}. Parameters are: $\Omega = 2.9\gamma$ $d = 0.05$, and $f^2 = 0.15$. They represent the realistic conditions at which the experiments of Sec.~\ref{sec:experiment} have been performed. } 
\label{fig:coefsreal}
\end{figure}

The most relevant effect cause of deviations in the RD measurement obtained in the laboratory lies in the possible spatial mode mismatch 
with the resonator eigenmode. 
Imperfect matching means that the field to be measured does not couple completely to the target optical resonance. The uncoupled fraction of light is reflected as if the resonator were a simple mirror, an effect that introduces an additional source of vacuum fluctuations in the quantum noise. It is not straightforward, however, to label such effect an `imperfection': although it slightly modifies the RD quantum measurement, it may actually improve the access to two-mode features in some situations, as we have observed in our experiment. We note that RD is able to surpass HD precisely because of vacuum modes added to certain spectral regions of the field; it is thus not surprising that adding vacuum modes in alternative ways could lead to beneficial results that break the symmetry between upper and lower sidebands in the way they contribute to the spectral quantum noise. For one thing, the spatial mismatch guarantees that there is always some light reaching the photodetector even at exact LO resonance ($\Delta=0$), a feature that avoids the problem of technical noise at this detuning region. Finally, we note that a longitudinally degenerate resonator (e.g. in confocal configuration) could effectively eliminate spatial mismatch effects on the quantum noise. 

To model such situation, one must consider that the spatial mode of the impinging beam finds a decomposition with at least two contributing modes in the spatial basis priviledged by the optical resonator. Writing the positive part $\hat E^{+}$ of the input electric field as~\cite{wolf} 
\begin{equation}
\hat E^{+}(t) = \vec F_1(\vec r)\;\hat A(t) + \vec F_2(\vec r)\;\hat B(t),
\end{equation}
where $\hat A(t)$ is the target resonator spatial mode (i.e. the mode to which we aim to perfectly couple the impinging beam) and $\hat B(t)$ is a contamination mode. The vectorial functions $\vec F_j(\vec r), j=1,2,\dots$, stand for the spatial profile of the electric field in certain basis modes (e.g. Hermite-Gaussian spatial modes). The photocurrent operator $\hat I(t)$ is proportional to the integral of $\hat E^{-}\cdot\hat E^{+}$ on the surface of the photodetector, where the functions $\vec F_j(\vec r)$ are assumed to respect orthonormality relations $\int \vec F_j(\vec r)\cdot\vec F_{j'}^*(\vec r) \mathrm{d}^2r=\delta_{jj'}$, yielding 
\begin{equation}
\hat I(t) = \hat A^\dag(t)\hat A(t) + \hat B^\dag(t)\hat B(t). 
\label{eq:photodetectionoperator}
\end{equation}
In this spatial modal basis, the input quantum state of LO mode appears as $\ket{LO} = \ket{\sqrt{1-f^2}\,\alpha}_1\ket{\sqrt{f}\,\alpha}_2$, where $f$ represents the fraction of modal contamination ($f=0$ for perfect spatial mode matching). Performing the quantum state average of the observable of Eq.~(\ref{eq:photodetectionoperator}) solely on the LO mode yields for the remaining spectral modes the photocurrent operator
\begin{align}
\hat I'(t) & \approx (1-f^2)\,|\alpha|^2 + f |\alpha|^2 \nonumber \\
& + \sqrt{1-f^2}\left(\alpha^* e^{i\omega_0 t}\,\hat A(t)+\alpha e^{-i\omega_0 t}\,\hat A^\dag(t)\right)\nonumber\\
& + f\left(\alpha^* e^{i\omega_0 t}\,\hat B(t)+\alpha e^{-i\omega_0 t}\,\hat B^\dag(t)\right),
\end{align}
where the prime superscript in $\hat I'(t)$ indicates that the quantum state average has already been performed on LO mode and only terms amplified by the LO have been kept. Disregarding the constant intensity contribution, the spectral photocurrent fluctuation is 
\begin{align}
\hat I_\Omega = & \sqrt{1-f^2}\left(\alpha^* \,\hat A_{\omega_0+\Omega}+\alpha \hat A_{\omega_0-\Omega}^\dag\right)\nonumber\\
& + f\left(\alpha^* \,\hat B_{\omega_0+\Omega}(t)+\alpha \,\hat B_{\omega_0-\Omega}^\dag\right).
\label{eq:Iomegacontaminabaseantiga}
\end{align}
Spatial mode mismatch implies that only the first term of Eq.~(\ref{eq:Iomegacontaminabaseantiga}) undergoes the modal transformation of Eq.~(\ref{eq:inputoutput}) upon interaction with the resonator. The second term is assumed off-resonant (i.e. the resonance frequency $\omega_c$ is assumed to be mode dependent). The spectral operator transformed by Eq.~(\ref{eq:inputoutput}) and normalized by the LO amplitude reads as 
\begin{align}
\hat J_\Omega(\Delta) = & \sqrt{1-f^2}|r(\Delta)|\left(R^*_\Omega(\Delta) \,\hat A_{\Omega} + R_{-\Omega}(\Delta)\, \hat A_{-\Omega}^\dag \right.\nonumber\\
&  + \left.\hat J_v\right)+ f\left(\hat B_{\Omega}(t)+\hat B_{-\Omega}^\dag\right),
\label{eq:Iomegacontaminabaseantigatransf}
\end{align}
where the notation has been simplified as in $\hat A_{\omega_0\pm\Omega}\rightarrow\hat A_{\pm\Omega}$. 
To obtain the RD transformation as measured by the photodetection, it is necessary to change the spatial modal basis from $\{\hat A,\hat B\}$ back to the detection basis $\hat a$, according to 
\begin{align}
\hat a_{\Omega} & = \sqrt{1-f^2}\hat A_\Omega + f\hat B_\Omega,\\
\hat c_{\Omega} & = -f\hat A_\Omega + \sqrt{1-f^2}\hat B_\Omega,
\end{align}
where the basis $\hat c_\Omega$ of the spatial mode (assumed in the vacuum state) orthogonal to $\hat a_{\Omega}$ is necessary to perform the inverse modal transformation in Eq.~(\ref{eq:Iomegacontaminabaseantigatransf}), yielding 
\begin{align}
\hat J_\Omega(\Delta) & = G^*_\Omega(\Delta)\;\hat a_\Omega + G_{-\Omega}(\Delta)\;\hat a_{-\Omega}^\dag +\hat J_v',
\label{eq:quantumobservable}
\end{align}
where 
\begin{align}
G_{\Omega}(\Delta) = (1-f^2)\,|r(\Delta)|\,R_\Omega(\Delta) + f^2
\end{align}
and the vacuum term is
\begin{align}
\label{eq:Jcontaminavacuumterm}
\hat J_v' = & f\sqrt{1-f^2}[\left(-|r(\Delta)|\,R^*_\Omega +1\right)\hat c_\Omega+ \\
& + \left(-|r(\Delta)|\,R_{-\Omega} +1\right)\hat c^\dag_{-\Omega}] + \sqrt{1-f^2}|r(\Delta)| \hat J_v. \nonumber
\end{align}
The observables of resonator detection with modal mismatch written in terms of spectral mode quadratures have the same form as in Eqs.~(\ref{eq:Jcosxy}) and~(\ref{eq:Jsinxy}) with the substitutions $x_{\Omega}\rightarrow x'_{\Omega}$ and $y_{\Omega}\rightarrow y'_{\Omega}$ defined as $G_{\Omega}(\Delta) = x_{\Omega}' + i y_{\Omega}'$. The vacuum terms are also substituted by the Hermitian and anti-Hermitian parts of Eq.~(\ref{eq:Jcontaminavacuumterm}). The main effect of modal contamination is to decrease the angular interval of phase space rotations, by removing redundant rotations in the ideal scenario of an impedance matched resonator ($d=0$). 
 For resonators in the intermediate scenario $0<d<1$, modal contamination can help access two-mode features with better sensitivity. 

\section{Quantum measurement of two-mode displaced quantum state}
\label{sec:experiment}

\subsection{Quantum states encoding phase information}

Displacement of the vacuum field is probably the simplest conceptual quantum operation capable of producing quantum states with clear phase information. Luckily, the experimental generation of coherent states in spectral field modes is equally simple and clear. 
Here we present the use of RD in a phase-modulated laser beam as a proof-of-principle demonstration of phase coherent detection. 

Let us then consider the \textit{`phase modulation'} of an intense field by an electro-optical modulator (EOM). In this process, the refractive index of the EOM crystal is modified by an external electric field, creating a controllable phase delay on the laser beam passing through the crystal. Therefore the classical electric field of light $E(t)=E_0 \exp\left[i(\omega t+\beta)\right]$ oscillating at optical frequencies can be periodically delayed by a radio-frequency electric field producing a phase shift $\beta \rightarrow \beta(t) = 2\beta_0\cos(\Omega t+\Phi)$. If the modulation amplitude is small enough ($\beta_0\ll2\pi$), the phase-modulated laser can be described by the electric field amplitude
\begin{align}
\label{eq:lasermodulation}
E(t) \approx & \;E_0\,e^{i\omega_0 t} \\
& +iE_0\beta_0\,e^{i\Phi}\,e^{i(\omega_0+\Omega) t} +iE_0\beta_0\,e^{-i\Phi}\,e^{i(\omega_0-\Omega) t},\nonumber
\end{align}
where higher frequency components can be disregarded for small modulation depth. 
From the point of view of the two-mode quantum mechanical field, Eq.~(\ref{eq:lasermodulation}) states that the semi-classical phase modulation corresponds to the displacement of upper and lower sideband modes with coherent states possessing the amplitudes $\alpha_\Omega = iE_0\beta_0\,e^{i\Phi}$ and $\alpha_{-\Omega} = iE_0\beta_0\,e^{-i\Phi}$. They represent complex conjugated numbers if the real and imaginary axes are interchanged (i.e. equivalent to a local rotation of field modes). The quantum state of sideband modes is Gaussian and separable, and equal to 
\begin{equation}
\ket{\psi} = \ket{\alpha_\Omega}_\Omega\otimes\ket{\alpha_{-\Omega}}_{-\Omega}.
\label{eq:quantumstatesidebandEOM}
\end{equation}
The motivation for the name `phase modulation' is made clear in the semi-classical interpretation of quantum noise~\cite{mandelScomplex,pra2013}. In this picture, valid as long as second-order moments (sufficient to describe Gaussian states) are concerned, `effective' quadrature operators can be found to succintly describe the photocurrent spectral noise power by using only half the number of actual field modes. The semi-classical amplitude and phase quadrature operators, respectively defined as $\hat{\mathcal{P}}_\Omega = \hat a_\Omega + \hat a^\dag_{-\Omega}$ and $\hat{\mathcal{Q}}_\Omega = -i(\hat a_\Omega - \hat a^\dag_{-\Omega})$,
yield over the quantum state of Eq.~(\ref{eq:quantumstatesidebandEOM}) the mean amplitudes 
\begin{equation}
\ave{\hat{\mathcal{P}}_\Omega}= 0
\quad\mathrm{and}\quad
\ave{\hat{\mathcal{Q}}_\Omega}= 2E_0\beta_0\,e^{i\Phi},
\label{eq:semiclassicalquadsdisplacement}
\end{equation}
i.e. only the semi-classical phase quadrature is displaced, thereby justifying the nomenclature. 
In terms of \textit{bona fide} quadrature observables of spectral modes, the semi-classical quadratures read as 
\begin{align}
\hat{\mathcal{P}}_\Omega = \hat p_s + i \hat q_a, \qquad
\hat{\mathcal{Q}}_\Omega = \hat q_s - i \hat p_a.
\end{align}
Hence Eq.~(\ref{eq:semiclassicalquadsdisplacement}) furnishes the quantum state averages of sideband modes quadrature observables as 
\begin{align}
{\textstyle\frac{1}{\sqrt2}}\ave{\hat p_{\Omega} +\hat p_{-\Omega}} = 0, & \qquad
{\textstyle\frac{1}{\sqrt2}}\ave{\hat q_{\Omega} +\hat q_{-\Omega}} = s\cos\Phi, \\
{\textstyle\frac{1}{\sqrt2}}\ave{\hat q_{\Omega} -\hat q_{-\Omega}} = 0, & \qquad
{\textstyle\frac{1}{\sqrt2}}\ave{\hat p_{\Omega} -\hat p_{-\Omega}} = s\sin\Phi,
\label{eq:phasemodulationtheory}
\end{align}
where $s = 2E_0\beta_0$. 

\begin{figure}[t]
\epsfig{width=0.9\columnwidth,file=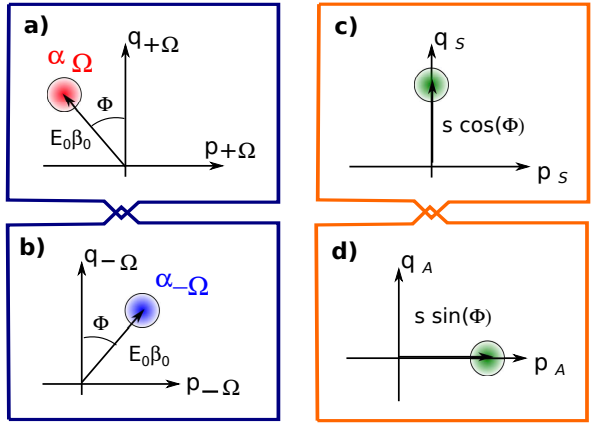}
\caption{Two-mode quantum state produced by laser phase modulation. Insets \textbf{a)} and \textbf{b)} depict the quantum state in the modal basis of sidebands. The quantum state is represented in the modal basis of symmetric $\mathcal{S}$ and antisymmetric $\mathcal{A}$ modal basis in insets \textbf{c)} and \textbf{b)} }  
\label{fig:quantumstate}
\end{figure}

For an ideal EOM, supposed not to include technical noise on the sideband modes, the conditions above correspond to the generation of two classically correlated coherent quantum states. One coherent state results from the displacement of the phase quadrature of $\mathcal{S}$ mode [Eq.~(\ref{eq:phasemodulationtheory})] and the other stems from the amplitude displacement of $\mathcal{A}$ mode [Eq.~(\ref{eq:phasemodulationtheory})]. On this modal basis, the phase $\Phi$ of phase modulation only determines the size of the displacements on fixed directions, since the quantum state then reads as
\begin{equation}
\ket{\psi} = \ket{\alpha_s}_s\otimes\ket{\alpha_a}_a,
\end{equation}
where $\alpha_s= is\cos\Phi$ and $\alpha_a = s\sin\Phi$.
On the modal basis of spectral sidebands, however, $\Phi$ controls the phase of displacement amplitudes in each spectral mode. In fact, the coherent state on mode $\pm\Omega$ has the amplitude $\alpha_{\pm\Omega}$ previously discussed, so that $\alpha_\Omega = - \alpha_{-\Omega}^*$. 
To access this phase information, RD requires the establishment of a well-defined relative phase between the generated spectral quantum state and the measured spectral photocurrent component. We achieve this regime by utilizing the same electronic reference to generate and measure the spectral sideband modes~\cite{naturebreitenbach}, a situation completely analogous to the usual practice of utilizing the same laser to produce the quantum state and homodyne it. 

In our experiment, the two-mode spectral quantum state is generated by an EOM (fed by the eLO) acting on a laser beam, in this manner performing the displacement quantum operation on both sideband modes. 
The input laser beam plays the role of the LO field with respect to which RD delays and attenuates the sideband modes, while the eLO becomes the electronic reference with respect to which spectral photocurrent components are defined by modulation. Since the phase of the spectral coherent states depend both on the LO and the eLO phases, we ensure good phase relations to exist in the two `downmixing' processes involved in resonator detection. Firstly, the optical downmixing between LO and sidebands quantum state defines the phase reference of amplitude and phase quadratures in phase space; secondly, the electronic downmixing between the photocurrent and the eLO allows us to define the cosine and sine spectral components, a step that is in general not pursued in experiments with quantum noise. Even though we perform two distinct steps of phase sensitive downmixing, we cannot claim that LO and eLO are in fact phase coherent to one another. The reason why the relative phase diffusion between LO and eLO factors out in our quantum measurement stems from the fact that the only phases that matter to our signal concern the quantum state: as long as it is coherent with both LO and eLO at the same time, the quantum \textit{measurement} may be constructed to occur in a phase coherent regime. 
This situation is in fact very common in experiments in quantum optics: by optical phase one usually means the phase of a light beam with respect to itself in a different time or position -- the absolute phase is inconsequential. Our setup makes the absoltute phases of both the optical LO and the eLO irrelevant at the same time, making it unnecessary to make them coherent to one another. 

With such arrangement, we are able to produce different quantum states by either changing the eLO amplitude or the relative phase between the fraction of eLO power fed to the EOM and that sent to the spectral analysis. In changing the phase, it is also correct to say that we produce the same quantum state, but vary the spectral component being measured (equivalent to varying $\Theta$). Here we adopt the first interpretation, so as to keep the quantum measurement fixed and the quantum state tunable.

\subsection{Experimental setup}

\begin{figure}[ht]
\epsfig{width=\columnwidth,file=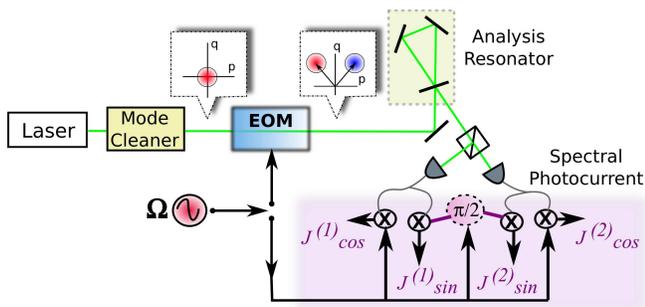}
\caption{Experimental setup. The laser beam is modulated by the EOM and coupled to an optical cavity to perform RD. The reflected beam is measured by two photodetectors providing the quantum noise and the SQL simultaneously. Spectral analysis of the photocurrent components is performed with the same electronic reference used to drive the EOM and produce the quantum state, in this manner achieving effective phase-locked detection. Sine and cosine photocurrent components are individually sampled and recorded by a commercial A/D converter board. } 
\label{fig:setup}
\end{figure}

The laser system is comprised of a frequency-doubled diode-pumped Nd:YAG laser at 532~nm (Innolight Diabolo) and spectrally filtered by an optical resonator to achieve shot noise limited spectral sideband modes (vacuum) at the intended analysis frequency as depicted in Figure~\ref{fig:setup}. 
 The sideband quantum state is produced by phase modulation of the laser beam with an EOM at $\Omega = 17$~MHz. Resonator detection is performed by employing an optical cavity with $5.9(3)$~MHz resonance bandwith [$\Omega/\gamma=2.9(2)$] and impedance matching parameter given by $d = 0.05$. Spatial mode matching achieves 86\% coupling ($f^2=0.15$) with the TEM00 Hermite-Gaussian mode. Although this value could be easily made very close to 100\% in our experiment (typically $>99.5\%$), lower values of spatial matching provides better access to two-mode features of the quantum state in our particular situation, as ractified by the model of Eq.~(\ref{eq:quantumobservable}). 
Fig.~\ref{fig:coefsreal} depicts the RD quadrature coefficients of Eq.~(\ref{eq:quantumobservable}) as they stand in our experiment.

Photodetection is realized by two amplified detectors with 25~MHz bandwidth. Each photodetector separates the photodiode photocurrent by frequency: the transmission of a low-pass filter with 10~kHz cut-off frequency samples the beam mean intensity (DC signal), while the high-frequency components from 10~kHz (HF signal) yield the quantum noise of interest. 
A single spectral photocurrent component is selected from the temporal signal by downmixing it with the use of an electronic local oscillator (eLO) with frequency $\Omega$ (fig. \ref{fig:setup}) and filtering the result in low-pass with 300~kHz cut-off frequency (or 600~kHz full width).
The two electronic downmixing components (cosine and sine, or, equivalently, in-phase and in-quadrature with respect to the eLO) of each detector are recorded by an analog-to-digital (A/D) converter connected to a computer at the acquisition rate of 600~kHz. 
The subtraction of demodulated HF components stemming from the two photodetectors provides the SQL, and their sum is further analyzed, giving rise to the spectral photocurrent components.
Quantum state reconstruction is realized by scanning the cavity length with a piezoelectric element holding one of the cavity mirrors. The resonator frequency $\omega_c$ is in this manner linearly scanned accross the spectral modes of interest. Each scan has duration of 0.75~s and collects 450,000 quantum measurements of each spectral photocurrent component.

\subsection{Experimental results}

\begin{figure}[ht]
\epsfig{width=0.8\columnwidth,file=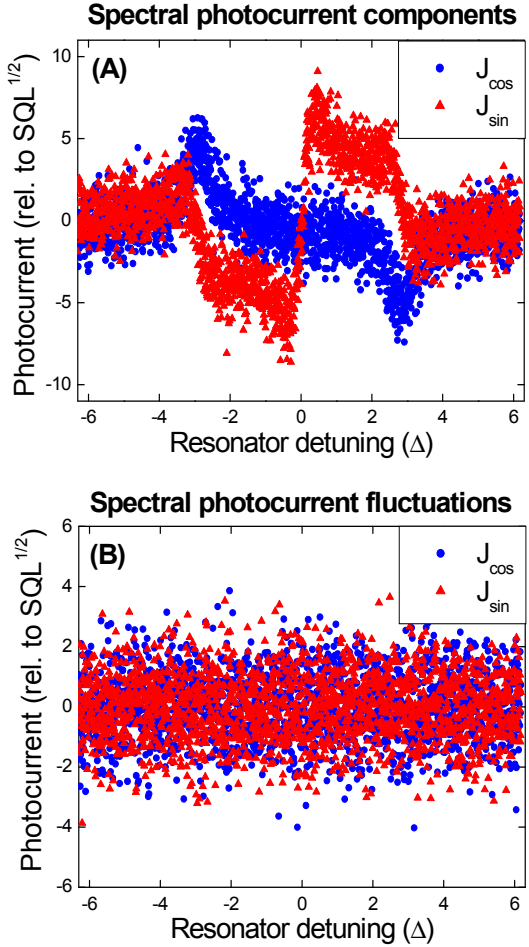}
\caption{Spectral components of the photocurrent as functions of resonator detuning $\Delta$. Each data point corresponds to the realization of a quantum measurement of either $\hat J_{\cos{}}$ (blue cicles) or $\hat J_{\sin{}}$ (red triangles). (A) Raw data normalized to the square root of the SQL. (B) Quantum fluctuations appearing on the data on top (i.e. mean values have been subtracted). Each figure shows a sample of 2,000 data points from the original 450,000 quantum measurements in each curve. } 
\label{fig:dados}
\end{figure}

Figure~\ref{fig:dados}(A) presents the series of individual RD quantum measurements (normalized to the SQL) of the phase modulated laser as function of resonator detuning $\Delta$. 
The phase coherent nature of the quantum measurement guarantees that the cosine ($\hat J_{\cos{}}$) and sine ($\hat J_{\cos{}}$) components are truly accessing certain marginal distributions of the two-mode quantum state Wigner function as the resonator detuning $\Delta$ is varied. It can be noticed that the signal shows not only the usual quantum fluctuations around the null value, but also the mean values of spectral components at each detuning. Fig.~\ref{fig:dados}(B) complements the picture by isolating the fluctuations of Fig.~\ref{fig:dados}(A) obtained by subtracting from each quantum measurement the mean value of 200 data points around it (high-pass filter). This step in the analysis is not necessary, but facilitates the separate visualization of first- and second-order photocurrent moments to be formally annalyzed later. 

\begin{figure}[ht]
\epsfig{width=0.8\columnwidth,file=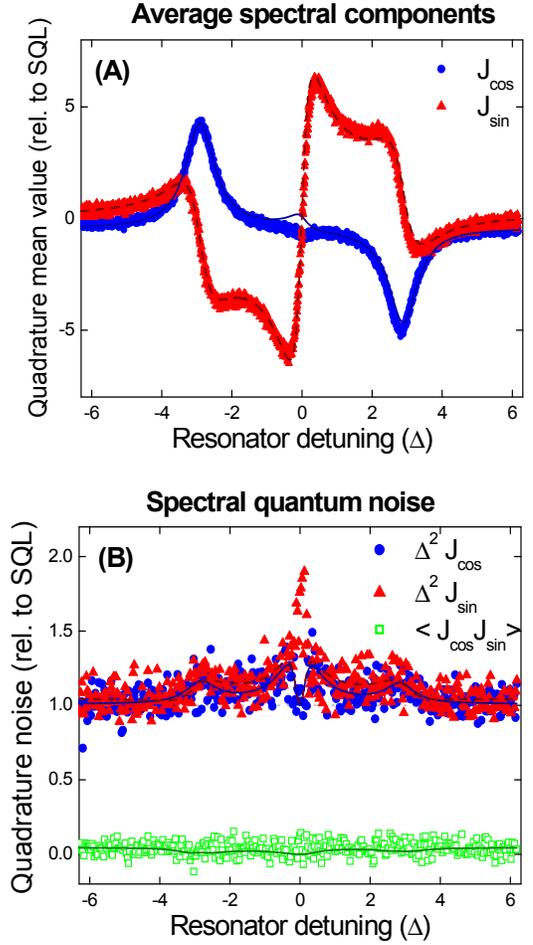}
\caption{Moments of the cosine and sine photocurrent components of Fig.~\ref{fig:dados} corresponding to the quantum measurement operators $\hat J_{\cos{}}$ and $\hat J_{\sin{}}$ [Eq.~(\ref{eq:quantumobservable})]. (A) First-order moments $\ave{\hat J_{\cos{}}}$ (blue circles) and $\ave{\hat J_{\sin{}}}$ (red triangles). (B) Second-order moments $\Delta^2\hat J_{\cos{}}$ (blue circles), $\Delta^2\hat J_{\sin{}}$ (red triangles), and $\ave{\hat J_{\cos{}}\hat J_{\sin{}}}$ (green squares). Curve fits involving the measurement operator model of Eq.~(\ref{eq:quantumobservable}) are depicted by solid and dashed lines on top of the respective data set.  }
\label{fig:ajustes}
\end{figure}

We extract information from the data in Fig.~\ref{fig:dados} in two steps. Firstly, we address the photocurrent mean values $\ave{\hat J_{\cos{}}}$ and $\ave{\hat J_{\sin{}}}$ calculated over 200 individual quantum measurements, yielding curves with 22,500 values of detuning. Those curves are presented in Fig.~\ref{fig:ajustes}(A) and provide information on the quadrature operator mean values $\ave{\hat p_{\pm\Omega}}$ and $\ave{\hat q_{\pm\Omega}}$ by fitting the quantum state average of the measurement operator of Eq.~(\ref{eq:quantumobservable}) to the data (solid and dashed lines). 
We note that the mean value resembles the `error signal' obtained in the Pound-Drever-Hall technique for resonator stabilization~\cite{PDH}. 
Secondly, we calculate from Fig.~\ref{fig:dados}(B) the second-order photocurrent moments -- variances $\Delta^2\hat J_{\cos{}}$ and $\Delta^2\hat J_{\sin{}}$ and the correlation $\ave{\hat J_{\cos{}}\hat J_{\sin{}}}$ -- to extract the covariance matrix of the two-mode Gaussian quantum state. The elements of the covariance matrix are obtained by fitting the quantum state average of the square of the measurement operator of Eq.~(\ref{eq:quantumobservable}), a result presented in Fig.~\ref{fig:ajustes}(B) by the solid and dashed lines. 

To perform the aforementioned RD model curve fits, we first obtain the general shapes of the curves $G_{\pm\Omega}$ as functions of $\Delta$ (depicted in Fig.~\ref{fig:coefsreal}) by simple analysis of the DC signal characterizing the resonator intensity reflection profile (a usual resonance curve, not shown). The DC signal provides the resonator parameter $d$ as well as the scaling factor that allows us to callibrate the detuning $\Delta$ relative to the resonator bandwidth. The quadrature operator moments appear in the data fitting as coefficients to the real and imaginary parts of the curves $G_{\pm\Omega}(\Delta)$. In this manner, the first-order moments [Fig.~\ref{fig:ajustes}(A)] of the measured photocurrent (Fig.~\ref{fig:dados}) can be understood (in the context of the curve fitting used to extract quadrature operator moments) as a sum of the curves depicted in Fig.~\ref{fig:coefsreal} weighted by the first-order moments of field quadrature operators. The same reasoning applies to the second-order moment curve fittings in Fig.~\ref{fig:ajustes}(B). 


Although in our experiment RD is not performed with a narrow resonator, since $\gamma\approx\Omega/3$, it is still possible to understand Fig.~\ref{fig:ajustes}(A) qualitatively. The detuning region near the LO resonance ($\Delta\approx0$) can be seen to reveal features in the $\mathcal{S}$/$\mathcal{A}$ modal basis by comparing the data profiles with the expected quantum state averages of Eq.~(\ref{eq:phasemodulationtheory}). In fact, the $\ave{\hat J_{\cos{}}}$ curve indicates that $\hat p_{s}\approx \hat q_s \approx 0$, while the curve for $\ave{\hat J_{\sin{}}}$ points at the existence of some displaced state in mode $\mathcal{A}$. Similarly, the detuning regions where the optical resonator is nearly resonant with one of the sidebands, at $\Delta\approx\pm\Omega/\gamma\approx \pm3$, indicate that both sideband modes $\pm\Omega$ are displaced by the same amount, since the curves are symmetric, essentially showing the same features for positive and negative detuning expected from Eq.~(\ref{eq:quantumstatesidebandEOM}). The quantitative analysis obtained by the theoretical curve fittings, represented by solid lines in Fig.~\ref{fig:ajustes}(A), attest that this is indeed the case. On the modal $\mathcal{S}$/$\mathcal{A}$ basis, we obtain the first-order quadrature operator moments $\ave{\hat p_s}= -0.6(7)$, $\ave{\hat q_s}= 2.2(5)$, $\ave{\hat p_{a}}= 11.8(7)$, and $\ave{\hat q_{a}}= 0.2(5)$. We note that these numbers are measured relatively to the scale determined by the SQL in phase space (i.e. the value 1 would indicate a coherent state displaced by the standard deviation of its Gaussian probability distribution). On the spectral basis of sideband modes, we obtain $\ave{\hat p_\Omega}= -8.8(7)$, $\ave{\hat q_\Omega}= 1.7(5)$, $\ave{\hat p_{-\Omega}}= 7.9(7)$, and $\ave{\hat q_{-\Omega}}= 1.4(5)$, whereby it is clear that sideband modes are displaced by equal amplitudes (given the experimental uncertainty) in conjugate directions in phase space, as expected by the quantum model of phase modulation. 

We note that the usual phase mixed detection would erase phase information encoded in the first order moments, completely nullifying them and artificially increasing the second-order moments to erroneously identify the phase modulated laser as possessing excess (semi-classical) phase noise. The two-mode sideband quantum state would then appear to show zero quadrature average and balanced excess noise: essentially, a thermal state \cite{prl2013}. Measurement mixedness would in this case be completely transferred to a perceived lack of purity of the quantum state (since the thermal state has the lowest degree of purity for a given temperature). Phase information allows us to perform a pure measurement and hence correctly identify the quantum state as very close to a coherent state in the sideband modes, indeed a very different situation from the inherent classical randomness of a thermal state.

Fig.~\ref{fig:ajustes}(B) depicts the experimental results regarding the photocurrent noise power and its interpretation in terms of the Gaussian quantum state covariance matrix. Three possible experimental combinations are possible: $\Delta^2\hat J_{\cos{}}=\ave{(\hat J_{\cos{}}-\ave{\hat J_{\cos{}}})^2}$, $\Delta^2\hat J_{\sin{}}=\ave{(\hat J_{\sin{}}-\ave{\hat J_{\sin{}}})^2}$, and $\ave{(\hat J_{\cos{}}-\ave{\hat J_{\cos{}}})(\hat J_{\sin{}}-\ave{\hat J_{\sin{}}})}$, respectively corresponding to the spectral noise power of photocurrent cosine and sine components, and to the correlation between those components. Variances and correlations are calculated over groups of 1,000 quantum measurements, and hence one resonator scan is composed of 450 detuning values. Theoretical curve fittings to the data, shown on top of the respective data set, involve the square of the measurement operator of Eq.~(\ref{eq:quantumobservable}). Similarly to the reasoning presented in the analysis of Fig.~\ref{fig:ajustes}(A), we may regard $\Delta \hat J_{\cos{}}$, $\Delta \hat J_{\sin{}}$ and $\ave{ \hat J_{\cos{}}\hat J_{\sin{}}}$ as sums of curves taken as functions of $\Delta$ weighted by elements of the covariance matrix. For instance, the coefficient of $\Delta^2 \hat p_{\Omega}$ is, according to Eq.~(\ref{eq:quantumobservable}), given by $x_\Omega^2(\Delta) = \mathrm{Re}\{G_\Omega^*\}^2$, a function of $\Delta$ given by the square of the black solid curve on the top row of Fig.~(\ref{fig:coefsreal}). In fact, the coefficients of the quadrature operator variances $\Delta^2 \hat p_{\pm\Omega}$ and $\Delta^2 \hat q_{\pm\Omega}$ are given by the square of the functions of $\Delta$ seen in Fig.~(\ref{fig:coefsreal}), respectively $x_{\pm\Omega}(\Delta)$ and $y_{\pm\Omega}(\Delta)$. Correlations between different quadrature operators contribute to the noise curves of Fig.~\ref{fig:ajustes}(B) as products of the respective functions of $\Delta$, as expected. Given the generally asymmetric shapes of those coefficients, it is clear that the data in Fig.~\ref{fig:ajustes}(B) favors symmetric noise in modes $\pm\Omega$. Furthermore, all noise terms are nearly shot-noise limited. The quantitative analysis performed by the curve fittings yield the spectral operator moments $\Delta^2\hat p_{\Omega} =\Delta^2\hat q_{-\Omega} = 1.25(3)$ and $\Delta^2\hat q_{\Omega} =\Delta^2\hat p_{-\Omega} = 1.28(3)$, showing that the sideband quantum states are not exactly coherent states, but rather present slight excess (classical) noise. These results indicate that the EOM introduces a small amount of balanced thermal noise in the sideband modes, 
probably due to Johnson noise in the driving electronics. According to the data fit, the energy imbalance is proportional to $\ave{\hat J_{\cos{}}\hat J_{\sin{}}} = (\Delta^2\hat p_{\Omega} +\Delta^2\hat q_{\Omega}) - (\Delta^2\hat p_{-\Omega}+\Delta^2\hat q_{-\Omega})=-0.01(3)$, hence compatible with zero. 
In the basis of $\mathcal{S}$ and $\mathcal{A}$ modes, the EOM produces classical noise in the quadratures $\Delta^2\hat q_+=1.50(3)$ and $\Delta^2\hat p_-=1.03(3)$. According to Eq.~(\ref{eq:phasemodulationtheory}), that would be interpreted in the semi-classical picture as a slight addition of `phase' noise to the laser beam. 

Putting together the first- and second-order moments obtained with RD, the experimental curve of Fig.~\ref{fig:dados}(A) can be understood as the two-mode phase space rotation of the coherent state displaced by the EOM `smeared' by (roughly) the shot noise inherent to the Heisenberg uncertainty principle.

\begin{figure}[ht]
\epsfig{width=0.8\columnwidth,file=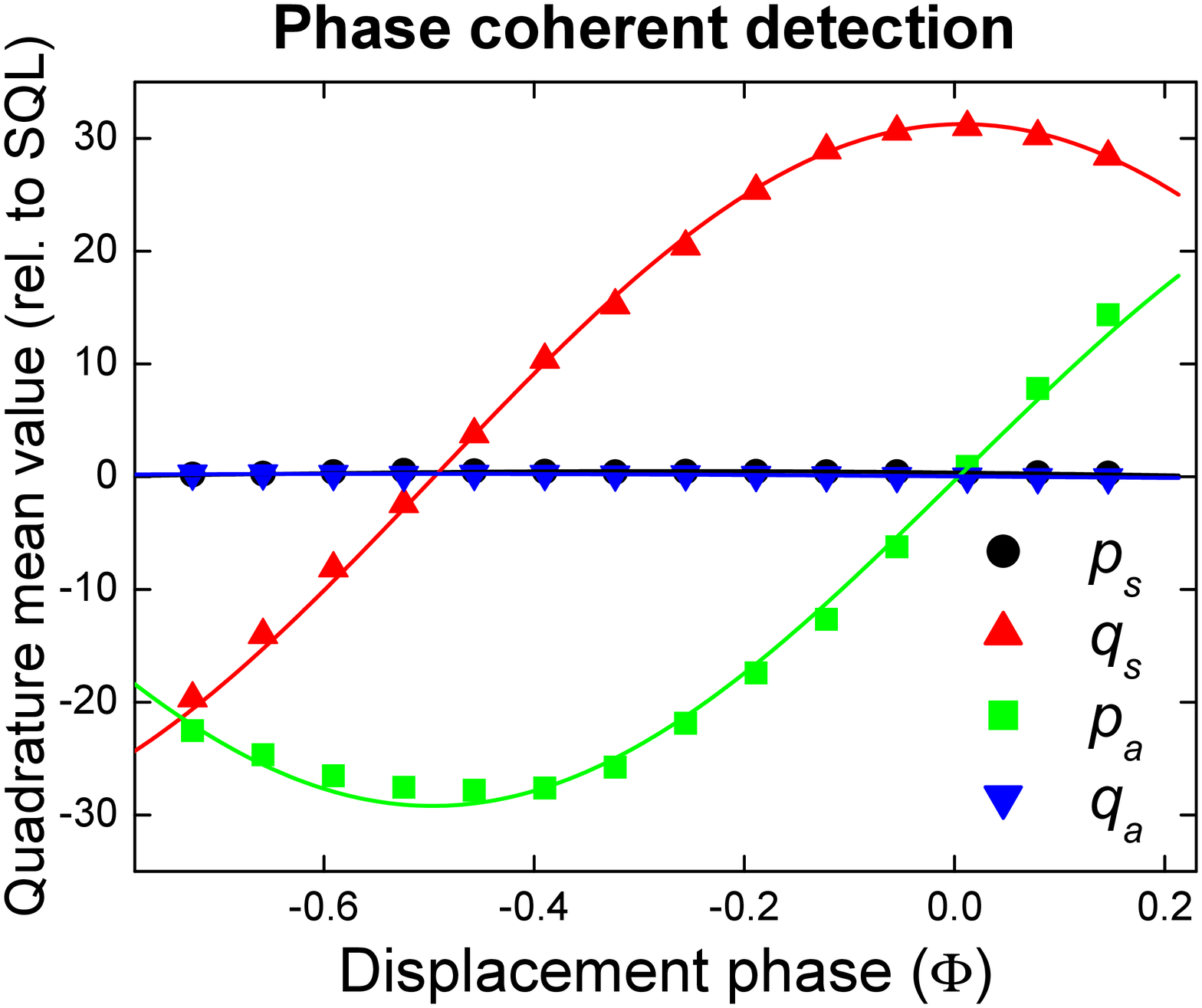}
\caption{Two-mode coherent state in modes $\mathcal{S}$ and $\mathcal{A}$ produced by varying the phase $\Phi$ of EOM modulation [Eq.~(\ref{eq:phasemodulationtheory})]. In particular, the quadratures $\hat p_s$ and $\hat q_a$ show null displacement for all values of $\Phi$.} 
\label{fig:phaselock}
\end{figure}

The ability to control the phase of the two-mode displaced quantum state is demonstrated in Fig.~\ref{fig:phaselock}. We have produced 14 different coherent quantum states by changing the displacement phase $\Phi$ fed to the EOM. For each value of $\Phi$, we obtain experimental curves analogous to those of Fig.~\ref{fig:ajustes}(A), to which we perform model fits to acquire the values of $\ave{\hat p_{\pm\Omega}}$ and $\ave{\hat q_{\pm\Omega}}$ or, by a change of modal basis, the moments $\ave{\hat p_s}$, $\ave{\hat q_s}$, $\ave{\hat p_a}$, and $\ave{\hat q_a}$. Fig.~\ref{fig:phaselock} presents the first-order moments in the modal basis $\mathcal{S}$ and $\mathcal{A}$ due to their increased simplicity [Eq.~(\ref{eq:phasemodulationtheory})] as the displacement phase $\Phi$ is varied. All quantum states are compatible with simple phase space rotations of the displacement values $\ave{\hat p_s}= -0.5(1)$, $\ave{\hat q_s}= 31.3(5)$, $\ave{\hat p_{a}}= 29.2(5)$, and $\ave{\hat q_{a}}= -0.1(3)$.


\section{Conclusion}
\label{sec:conclusion}

The path to extend the tools of CV quantum state reconstruction to general unknown spectral quantum states~\cite{raymertomoPRL93,lvovsky09}, even those presenting non-Gaussian features, crucially depends on the experimental capability of achieving phase coherence in the measurement of quantum noise. 
Measurement of the field quadratures relies on the interference (or `optical downmixing') between the field of interest and a relatively intense `classical' field taken as the phase reference, the local oscillator (LO). HD techniques both in the time and in the spectral domains, as well as RD, are based on this principle. In the spectral domain, however, an additional step is required to select a given frequency component of the photocurrent quantum fluctuations, performed by mixing it with an external electronic local oscillator (eLO). This process accesses the quantum state of two spectral modes equally separated from the LO optical frequency, the spectral sideband modes. 
Although the LO used in the first step is normally also involved in the generation of the quantum states of interest, in this manner ensuring phase coherence between them~\cite{coherencefiction}, the same does not hold true for the eLO, typically taken as an independent electronic source completely unrelated to the remaining quantum dynamics: in the current state of affairs, the interpretation of the spectral quantum noise as a pure quantum measurement is still a `convenient fiction' with limited context of validity (i.e. symmetric and Gaussian quantum states~\cite{gauss2015}) and thus insufficient to perform complete quantum state reconstruction free of prior knowledge. Ultimately, the mixedness of the measurement operator stems from the incoherent character of the electronic downmixing process.

The measurement technique of RD grants access to `hidden' sectors of the two-mode phase space of the spectral quantum state by providing modal-dependent attenuation~\cite{ralphSidebandsSeparationPRA05} and phase delay with the aid of a controllable optical resonance, even in the usual situation of a phase-mixed measurement~\cite{prl2013}. In order to obtain a pure measurement operator, phase coherent detection of the spectral photocurrent must be performed. We have shown that RD becomes in this case also a \textit{complete} two-mode measurement technique, providing access to any direction of observation in the four-dimensional phase space where the Wigner function describes the quantum state -- a fundamental condition to formally perform quantum state reconstruction of spectral modes. Phase coherent detection allows one to formally associate the photocurrent statistics with the probability distributions of \textit{bona fide} modal quadrature observables, in this manner bringing the prospect of assumption-free quantum state reconstruction~\cite{ralphSinglePhotonSidebandsPRA08}. 

Phase coherent detection requires the existence of good phase relation between the quantum state and the local oscillators used as references (both in the optical and electronic downmixing processes). We demonstrate phase coherent RD by measuring a simple quantum state possessing the desired phase information, a displaced two-mode quantum state. We keep track of the phase coherence between the quantum state and the quantum observable by employing the laser beam and the EOM electronic modulation signal as simultaneous references~\cite{naturebreitenbach}. In this manner, we are able to recover the displacement phases of the two-mode quantum state without needing to resort to optical phase locking techniques. By coherently measuring the two (cosine and sine) spectral photocurrent components, we demonstrate in experiment the capability to access the complete two-mode quantum state in phase space. The technique in principle works for any quantum state, even those presenting non-Gaussian statistics. 

In most experimental situations producing quantum states of the field, a weak seed beam, 
generated
 by an EOM, should suffice to introduce the necessary phase reference in the quantum state in order to later recover it in the measurement process~\cite{naturebreitenbach}. Alternatively, one could employ as optical LO laser beams showing spectral linewidth compatible with the inverse of the time needed to perform the full tomographic reconstruction of the quantum state~\cite{subhertzlaser}. Those procedures would introduce the missing degree of rigor in experiments aiming to arbitrarily manipulate the quadrature observables of spectral modes of light. Using the electronic seed as a `reference to itself' in the measurement process
 in the spectral domain is akin to the usual procedure of employing the laser beam as the `optical reference oscillator to itself' in order to keep track of the optical phase reference. 
 As we show here, the same care must be exercised when performing the spectral analysis of the photocurrent fluctuations: leaving the spectral phase free leads to the onset of mixed quadrature measurements, a clear limitation for the implementation of quantum information protocols requiring formal pure measurement operators and quantum feedback. Resonator detection adds to those capabilities by offering a complete measurement of the two-mode spectral field quantum state in a novel phase coherent regime.

\acknowledgments

This work was supported by grants \#2010/52282-1, \#2010/08448-2, \#2009/52157-5, S\~ao Paulo Research Foundation (FAPESP), CNPq, and CAPES (PROCAD program). This research was performed within the framework of the Brazilian National Institute for Science and Technology in Quantum Information (INCT-IQ).

\end{document}